
\def\frac#1#2{{#1\over#2}}

\magnification=\magstep1   \openup1\jot  \vsize=9truein
\font\titlefont=cmcsc10
scaled \magstep2

\hskip 4in DAMTP-R/95/4

\hskip 4in gr-qc/9502017

\hskip 4in January, 1995

\vskip 1in

\centerline {\titlefont Quantum Coherence and Closed Timelike Curves  }  \vskip
1truein  \centerline {S.~W.~Hawking} \vskip .5truein  \centerline {Department
of
Applied Mathematics and Theoretical Physics}  \centerline {University of
Cambridge}  \centerline {Silver Street}  \centerline {Cambridge CB3 9EW}
\centerline {UK} \vskip 1truein  \centerline {\bf Abstract}
 \bigskip

Various calculations of the $S$ matrix have shown that it seems to be non
unitary
 for interacting fields when there are closed timelike curves. It is argued
that
 this is because there is loss of quantum coherence caused by the fact that
part
of the quantum state circulates on the closed timelike curves and is not
measured at infinity.
 A prescription is given for calculating the superscattering matrix $\$ $ on
space times whose parameters can be analytically continued to obtain a
Euclidean
metric.
 It is illustrated by a discussion of a spacetime in with two disks in flat
space are  identified. If the disks have an imaginary time separation, this
corresponds to a heat  bath. An external field interacting with the heat bath
will lose quantum coherence.  One can then analytically continue to an almost
real separation of the disks. This  will give closed timelike curves but one
will still get loss of quantum coherence.

\vfil \eject

\beginsection 1. Introduction

 This paper is about what sense, if any, can be made of quantum field theory on
a  spacetime background that contains closed timelike curves. The development
of
causality  violation in a bounded region is classically forbidden in the sense
that it can not  occur if the weak energy condition holds [1]. However, quantum
field theory in curved  spacetime  has many examples like the Casimir effect
where the expectation value of  the energy momentum tensor fails to obey the
weak energy condition. It has therefore  been suggested [2] that an advanced
civilization might be able to create a wormhole in  spacetime which could be
used to travel into the past. This has led  to a lot of  interest in the
problem
of the formulation and behavior of quantum field theory in  spacetimes with
closed timelike curves.

In general, it seems that divergences in the energy momentum tensor occur when
one  has closed or self intersecting null geodesics [3]. These divergences may
create  spacetime singularities which  prevent one from traveling through to
the
region of  closed timelike curves [1]. However, Kim and Thorne [3] have
suggested that quantum  gravitational effects may smear out the divergences and
lead to a non singular  spacetime. It is therefore of interest to consider the
properties of quantum field  theory in spacetimes with closed timelike curves.

In particular, a number of authors have studied what I shall call, confined
causality  violating spacetimes. In these, there are well behaved initial and
final regions and
 the causality violations are restricted to a region in the middle. One can
make
this
 definition more precise but I shall refrain from doing so in order that I can
discuss
 as  wide a class of examples as possible. On such spacetimes, one might hope
to
calculate an $S$ matrix which would relate the quantum state in the final
region
to the  state in the initial region. Some authors have claimed [4, 5] that
this $S$ matrix  will be unitary for free fields but will be non unitary if
there are interactions. To
 try to make sense of such non unitarity, Hartle [6] has suggested that the
usual  amplitudes should be normalised by a factor that depends on the initial
state. This  would restore  conservation of probability  but at the heavy price
of making quantum mechanics non  linear.  In principle,  one would be able
detect the non linearity produced by  a wormhole, which  an advanced
civilization might create in the distant future. There  would
 be a paradox, if this information were to cause the advanced civilization to
change  its  mind about creating the wormhole. But such paradoxes occur anyway
with closed  timelike curves.

Anderson [7] has suggested that one should evolve the quantum state only with
the
unitary part of the $S$  matrix $U=(SS^{\dagger})^{-1/2}S$. The trouble with
this proposal is that it doesn't obey the usual composition law [8]: the
unitary
part of $S_1 S_2$ is not the product of the unitary part of $S_1$ and $S_2$
separately. A third proposal is to extend the $S$ to be a unitary
transformation
on a larger Hilbert space [8]. The trouble with this idea is that the larger
Hilbert space may have to have an indefinite metric.

The message of this paper is that there is no need to propose non linear
modifications  of
 quantum theory, or indefinate metrics on Hilbert space. The reason that the
$S$
matrix, calculated according to the usual  rules,
 is non unitary, is that there is loss of quantum coherence when there are
closed  timelike curves. This means that the probability to go from an initial
state to a  final state is given by a superscattering operator, $\$$, rather
than by $SS^{\dagger}$.
 Thus it does not matter that the object that one might think was the $S$
matrix, is  not unitary.

A proposal has been made by Deutsch [9] and Politzer [10] for calculating the
evolution  in the presence of closed timelike curves. This approach is based on
finding a  consistent  solution for the density matrix. This solution will
involve loss of quantum coherence
 in general.  However, it will also depend in a non linear way on the initial
state [11],  which means that one loses the superposition principle.

If one simply requires that the quantum theory is linear, the most general
relation  between the initial and final situations is not an $S$ matrix but a
superscattering  operator, $\$$, that maps initial density matrices, to final
ones. In what follows  it will be helpful to use index notation. I shall
represent a vector in a Hilbert  space, by a quantity with an upper index.
$$\lambda^A \in {\cal H} $$ The corresponding vector in the complex conjugate
Hilbert space, will carry a lower  index.
$$\overline{\lambda}_A\in\overline{\cal H} $$ The $S$ matrix is a linear map
from the initial Hilbert space to the final Hilbert  space,
 so it can be written as a two index tensor. $$\psi_+{}^A=S^A{}_B\psi_-{}^B $$
However, the most general description of the quantum state of a system is not a
vector  in  a Hilbert space, but a density matrix. This can be regarded  as a
Hermitian two index  tensor on Hilbert space. Then the most general linear
evolution is given by a  four index tensor that maps initial density matrices
to
final ones.  $$\rho_+^A{}_B=\$^A{}_{BC}{}^D\rho_-{}^C{}_D $$

Many familiar quantum systems  obey the Axiom of Asymptotic completeness [12].
This  requires  that the Hilbert space of the interaction region in the middle
is isomorphic to the  initial  and final Hilbert spaces. In other words, there
are unitary maps between the
 interaction region Hilbert space and the initial and final Hilbert spaces. If
this
 is the  case, there will evidently be a unitary map from the initial Hilbert
space to the final  Hilbert space. The superscattering operator $\$ $ will
factorize into the product  of an $S$ matrix and its adjoint.
$$\$^A{}_{BC}{}^D=S^A{}_C \overline{S}_B{}^D $$ In this situation, a density
matrix corresponding to a pure quantum state will be  carried  into a pure
quantum state. There will be no loss of quantum coherence.

However, there are quantum systems that do not obey the Axiom of Asymptotic
Completeness.
 An example is provided by a particle interacting with a heat bath. A heat bath
is not  in
 a single quantum state. Rather it can be in any quantum state $| n\rangle $
 with probability $\exp -{E \sb n\over T}$ In other words, it is in a mixed
quantum  state . A particle that is initially in a pure quantum state, which
interacts with the heat  bath,
 will end up in a mixed quantum state. This loss of quantum coherence is to
 be expected:  information about the original quantum state of the particle is
lost  into  the heat bath. However, I will give examples of systems with closed
timelike curves  that  are very similar to particles interacting with a heat
bath. The only difference is  that the temperature of the heat bath is
imaginary. This corresponds to a spacetime
  that is identified periodically in real Lorentzian time, rather than
periodically in  imaginary time, as in a normal heat bath. It should therefore
come as no surprise  that one gets loss of quantum coherence in these cases.
More generally, whenever one  has  confined causality violations, one has part
of the quantum state that is  circulating  on closed timelike curves. When one
makes measurements at infinity, one does not see  this part of the state. One
will therefore have to describe the state at infinity  by a mixed state,
obtained by tracing out over the part of the state that one can't see.

In this paper   I shall show that the usual rules of quantum theory seem to
lead
to  loss  of quantum coherence when there are closed timelike curves. I should
emphasize that  this  loss is not an optional feature that one can choose
whether or not to have in the  theory. Rather, like radiation from black holes,
it is an inevitable consequence of  the standard assumptions of quantum field
theory in curved spacetime. The only way to  protect the purity of quantum
states is either to abandon one or more of these  standard assumptions, or
subscribe to the Chronology Protection Conjecture [1]. This  says  that quantum
effects become so large when closed timelike curves are about to  appear
 that they either prevent the curves appearing or they bring the spacetime to
an
end
 at a singularity. In either case the laws of physics conspire to prevent
causality  violations.

\beginsection 2. Euclidean Approach

It is clear how to define quantum field theory on a curved spacetime background
that  is globally hyperbolic. That is to say,  it can be covered by a family of
Cauchy  surfaces. In this case, one can choose the commutator of two free field
operators  to be the half advanced minus half retarded Green function, which is
well defined.

$$[\phi (x), \phi (y)]=i D(x,y)$$ However, it is much less clear how to proceed
if the spacetime is not globally  hyperbolic, and in particular, if it contains
closed timelike curves. In this case,  it is not clear how to generalize the
half advanced minus half retarded Green function  and field operators at points
that are locally spatially separated may not  commute  because the points can
be
joined by a timelike curve that goes round a large loop  and returns to the
same
neighbourhood.

The approach I shall adopt is to analytically continue the parameters of the
spacetime
 with closed timelike curves to get a metric with a real Euclidean section. On
this  section, all the field operators commute and the Green functions are well
defined.  One then analytically continues back, both in the parameters of the
metric, and in the
 points themselves in a certain order, to get Green functions in the original
Lorentzian spacetime. One can then calculate the superscattering operator from
the  Green fun ctions according to certain rules which involve displacing the
points slightly into
 the complex. This is equivalent to the usual $i\epsilon $ prescription.

One can illustrate this with a simple causality violating spacetime which has
been  discussed by Politzer [8]. Two flat spacelike three dimensional disks of
radius  $R$  are located in Minkowski space at $t=-{1\over 2} T$ and $t={1\over
2}T$  and  the same spatial coordinates. One makes the following
identifications:

\item 1 The lower surface of the bottom disk is identified with the upper
surface  of the top disk.

\item 2 The upper surface of the bottom disk is identified with the lower
surface  of the top disk.

The resultant spacetime is geodesically incomplete at the edges of the disks.
However, one can impose boundary conditions there that make the wave equation
 well behaved, at any rate locally. The first identification is not very
significant and is imposed  just to avoid free surfaces. But the second
introduces closed timelike curves  in the region between the disks.

In order to define the Green functions, I shall first take the separation
between
 the disks $T$ to be pure imaginary. One then has a Euclidean spacetime with
time
 coordinate $\tau = it $.  On this one can define Green functions in the normal
way  as $n$ point expectation values.
$$G(x_1,x_2,\dots,x_n)=\langle\phi(x_1)\phi(x_2)\dots\phi(x_n)\rangle$$ $$=\int
 d[\phi] \phi(x_1)\phi(x_2)\dots\phi(x_n)e^{-I} $$

In the case of free fields, the $n$ point function will be built out of all
combinations of  two point functions. But if there are interactions, one will
have the usual Feynman diagram expansion in terms of the two point function  on
the background.

On the Euclidean spacetime, all points are spacelike separated from each other.
This means that the field operators at different points  commute with each
other.  Thus the $n$ point Green function does not depend on the order in which
the $n$  points are taken, as is obvious from the representation of the Green
function by a path  integral. On the other hand, the $n$ point Wightman
functions in Lorentzian  spacetime certainly do depend on the order of the
points, because the field  operators do not commute at
 timelike separated points. The way this comes about is that one analytically
continues the $n$ point expectation values from the Euclidean to the Lorentzian
regime, keeping a small  imaginary time displacement between each field point
 [12]. The displacement  is in the positive imaginary time direction between
each point in the Lorentzian
 expectation value reading from left to right. That is, for the expectation
value $$\langle \phi (x_1)\phi (x_2)... \phi (x_n)\rangle $$
 one requires that ${\rm Im}(x^0_1)<{\rm Im}(x^0_2)<\dots<{\rm Im}(x^0_n) $.
The purpose of the displacement is to evaluate the analytically continued
expectation values on the right side of the singularities that occur on the
 complex light cone. It is equivalent to the usual $i \epsilon $ prescription,
for
 integrating round the singularities in the propagator.

\beginsection 3. The Superscattering Operator $\$ $

I now come to the rules for calculating the superscattering matrix from  the
Lorentzian expectation values [12]. One makes the usual assumption,  that the
field in the initial and final regions can be expanded in terms  of
annihilation
and creation operators. $$\phi=\sum_i \lbrace f_i a_i + \overline{f}_i
a^\dagger_i \rbrace $$ The annihilation operators $a_i$ are multiplied by
positive frequency wave  functions $f_i$ and the creation operators
$a^{\dagger}$ are multiplied by
 negative frequency wave functions $\bar f_i$. One can invert this relation
 and express the annihilation and creation operators as integrals over
spacelike
surfaces of the field operator $ \phi $ with negative or positive frequency
wave
functions. $$a_i = \int_\Sigma
\overline{f}_i(x){\buildrel\leftrightarrow\over\nabla}_\mu  \phi(x) d\Sigma^\mu
(x) $$

One is interested in the expectation value of certain operators $Q$ in a given
initial state with density matrix $| \psi _-\rangle \langle \psi _-|$. This
will be given by $$\langle I ^{\dagger} Q I \rangle $$ where $I $ is a string
of
creation operators that create the given initial  state: $$ | \psi _-\rangle =
I
| 0_-\rangle $$
 In particular, one is interested in the probability of the final state
containing a set of particles created by the string $F$ of creation operators.
This would correspond to taking $Q=FF^{\dagger}$. Thus  the superscattering
matrix element between the initial state $| \psi _-\rangle \langle \psi _-| $
and  the final state $| \psi  _+\rangle \langle \psi _+| $ is determined by
$\langle I^\dagger FF^\dagger I \rangle$.

One can express the annihilation and creation operators in terms of the field
operators. In this way, the superscattering matrix can be calculated from an
integral of expectation values with initial and final wave functions. In order
to  get the right operator  ordering in the expectation value, these integrals
over the initial and final
 surfaces have to be slightly displaced in the imaginary time direction. The
rule is, the initial creation operators have the greatest displacement in the
 positive imaginary time
 direction. They are followed by the final annihilation operators, the final
creation operators, and then the initial annihilation operators. This is
illustrated in Figure 1.

In Minkowski space positive frequencies propagate only towards the negative
imaginary time direction, and negative frequencies propagate only towards
positive  imaginary time. This means that the data from the string $F
^{\dagger}$ that  corresponds to the  annihilation operators for the final
state
can propagate only upwards in imaginary  time. Because the final state
annihilation operators act on a space like surface  slightly above the real
time
axis, the only surface they can propagate to is the surface on
 which the initial state creation operators act. Similarly, the positive
frequency  data from the final state creation operators can only propagate
downwards,  and the only surface it can reach is that on which the initial
state
annihilation
 operators act.  Thus in this case, the diagram that represents going from
initial state $| \psi  _-\rangle \langle \psi _-| $ to the final state $| \psi
_+\rangle \langle \psi  _+| $ falls into two disconnected parts. This means
that
the probability for going  from initial  to final factors into an $S$ matrix,
corresponding to the upper part of the diagram,
 and its adjoint, corresponding to the lower part. In this situation, there is
no  loss of quantum coherence and the $S$ matrix is unitary.

Suppose however one identifies disks in Minkowski space at $\pm T $ where $ T $
is
 imaginary. Then negative frequency data from the final state annihilation
operators  will be able to propagate upwards in imaginary time to the upper
disk
and re-emerge
 at the lower disk. From there it can propagate upwards to the final state
creation
 operators, rather than the initial state creation operators. Similarly, the
data from the initial state creation operators can propagate downwards to the
initial
 state
 annihilation operators. In this way, one gets a diagram that is not divided
into two  parts by the real time axis. This means that the probability will not
factor  into an $S$ matrix and its adjoint, and there will be loss of quantum
coherence.  Physically, this is  what one would expect. The region between the
disks is identified periodically in
 imaginary time. Thus it corresponds to a heat bath and will be in a mixed
quantum  state. A free field would propagate straight through the heat bath and
not notice its  existence but an interacting field will be affected and will
lose quantum coherence to  the heat bath.

In the two disk example that has been considered, the expectation value of
 $FF^{\dagger}$ will be non zero even if the string $I $ of initial creation
operators is empty. This means one can detect particles in the final state even
 when there were none present originally. The reason one gets such energy non
conservation is that  one is considering field theory on a fixed background
that
is not time  translation invariant. One would expect energy conservation only
in
a full  quantum theory of gravity in which  one summed over all metrics as well
as all fields in those metrics. Nevertheless,
 in the
 context of quantum field theory in curved backgrounds, it may make sense to
consider the change in the final state brought about by the application of the
initial state creation and annihilation operators. Thus one should calculate
the
superscattering
 matrix not from $$\langle I^{\dagger} FF^{\dagger} I \rangle $$ But from
$$\langle I^{\dagger} FF^{\dagger} I \rangle - \langle FF^{\dagger}  \rangle $$

The idea is now to rotate the separation of the disks from imaginary to almost
real.
 One should keep a small imaginary part to the separation. This damps possible
divergences from high frequency modes by reducing them by a thermal factor,
$\exp  (-E/T)$, with $1/T$ imaginary but with a small positive real part.
Keeping a small  imaginary  part also means that one can rotate the separation
of the disks from imaginary to  almost real time without encountering any
singularities. If one didn't keep a small  imaginary part to the separation,
one
couldn't use analytical continuation from a  Euclidean  spacetime, to determine
the Green function, but would have to find some other  prescription.

There is a question of which direction one should rotate the separation from
imaginary
 time to almost real. In a path integral over metrics, presumably both
directions will  occur. I therefore think one should take the sum of the
rotations in both directions.
 This will ensure that the probabilities of going from initial to final are
real.

As long as one keeps a small imaginary part to the separation, one will lose
quantum  coherence, like in the pure imaginary separation case. Thus it seems
that there will  be loss of quantum coherence in the Lorentzian case.
Physically
this is reasonable,
 because one has external fields interacting with a heat bath at an imaginary
temperature
 with a small real part. More generally, one might expect loss of quantum
coherence  whenever one has closed timelike curves, because there will be a
part
of the quantum  state that one doesn't measure initially or finally.

The superscattering matrix for this spacetime will not conserve  energy because
one has  been considering quantum field theory on a fixed background and not
taking back reaction
 into account. In two dimensional black hole calculations, where we know how to
 include back reaction, one finds that the superscattering matrix conserves
energy [13].

\beginsection 4. Conclusion

I have shown that the reason the usual rules seem to lead to an $S$ matrix that
is
 non unitary is that there is part of the quantum state that circulates on the
closed
 timelike curves and is not measured at infinity. This leads to loss of quantum
coherence and a superscattering matrix that does not factor. Thus if one
multiplies the
 object that would normally be the $S$ matrix by its adjoint, one does not get
the  probability for going from the initial, to the final state. This means
there is no  reason for the $S$ matrix to be unitary.

My approach to calculating the superscattering operator in the presence of
closed
 timelike curves, consisted of two elements. The first was a set of rules that
give the  superscattering matrix in terms of the ordered Lorentzian expectation
values,  analytically continued to a neighbourhood of the real time axis. These
rules are not a  matter  of choice. They are forced on us by the usual
assumptions of quantum field theory on a  fixed background. The second element
of my approach, was to analytically continue
 the parameters of the Lorentzian metric, to obtain a Euclidean one. One could
uniquely
 define the Green functions in this metric to be the $n$ point expectation
values. One  could then analytically continue in both the field points and the
parameters of the  solution to get the Lorentzian expectation values, which
could then be used to  calculate the superscattering matrix. There might be
alternative, inequivalent ways
 of defining the Lorentzian expectation values  but for the reasons I have
given, I  think  that any reasonable alternative would also give loss of
quantum
coherence.

Spacetimes with closed timelike curves show that loss of quantum coherence is
not  confined to black holes, but can occur with other spacetimes with non
trivial causal  structure. This is important, because some people have claimed
that the Planck scale  physics at the end of black hole evaporation, will
restore quantum coherence. However,  in a causality violating spacetime, the
curvature could be small everywhere and Planck  scale physics would not come
in.
This reinforces my conviction that quantum coherence
 really {\it is} lost in black hole evaporation.

Personally, I don't believe that closed timelike curves will occur, at least on
a  macroscopic scale. I think that the Chronology Protection Conjecture will
hold and
 that divergences in the energy momentum tensor will create singularities
before
closed time like curves appear. However, if quantum gravitational effects
somehow  cut off these divergences, I'm quite sure that quantum field theory on
such a  background will show loss of quantum coherence. So even if people come
back from  the future, we won't be
 able to predict what they will do.

\beginsection  Acknowledgements

I am grateful to Mike Cassidy, Jim Hartle and David Politzer for discussions.

\beginsection References \baselineskip=20pt

[1] S. W. Hawking, Phys. Rev. {\bf D46}, 603 (1992)

[2] M. S. Morris, K. S. Thorne and U. Yurtsever, Phys. Rev. Lett. {\bf 61},
1446
(1988)

[3] S--W. Kim and K. S. Thorne, Phys. Rev. {\bf D43}, 3929 (1991)

[4] D. G. Boulware, Phys. Rev. {\bf D46}, 4421 (1992)

[5] J. L. Friedman, N. J. Papastamatiou and J. Z. Simon, Phys. Rev. {\bf D46},
4456 (1992)

[6] J. B. Hartle, Phys. Rev. {\bf D49}, 6543 (1994)

[7] A. Anderson, gr-qc 9405058 (1994)

[8] C. J. Fewster and C. G. Wells, hep-th 9409156 (1994)

[9] D. Deutsch, Phys. Rev. {\bf D44}, 3197 (1990)

[10] H. D. Politzer, Phys. Rev. {\bf D49}, 3981 (1994)

[11] M. J. Cassidy, gr-qc 9409003, {\it to be published} (1994)

[12] S. W. Hawking, Commun. Math. Phys. {\bf 87}, 395 (1982)

[13] S. W. Hawking, Phys. Rev. {\bf D50}, 3982 (1994) \end